\documentclass[reprint,amsmath,amssymb,aps]{revtex4-2}

\usepackage{graphicx}
\usepackage{dcolumn}
\usepackage{bm}
\usepackage{xcolor}

\begin{document}


\title{Generating ultra-dense pair beams using 400 GeV/c protons}

\author{C. D. Arrowsmith$^1$}
\author{N. Shukla$^2$}
\author{N. Charitonidis$^3$}
\author{R. Boni$^4$}
\author{H. Chen$^5$}
\author{T. Davenne$^6$}
\author{D. H. Froula$^4$}
\author{B. T. Huffman$^1$}
\author{Y. Kadi$^3$}
\author{B. Reville$^7$}
\author{S. Richardson$^8$}
\author{S. Sarkar$^1$}
\author{J. L. Shaw$^4$}
\author{L. O. Silva$^2$}
\author{R. M. G. M. Trines$^6$}
\author{R. Bingham$^{6,9}$}
\author{G. Gregori$^1$ \vspace{0.1cm}}

\affiliation{$^1$Department of Physics, University of Oxford, Parks Road, Oxford OX1 3PU, UK}
\affiliation{$^2$GoLP/Instituto de Plasmas e Fusão Nuclear, Instituto Superior Técnico, Universidade de Lisboa, 1049-001 Lisboa, Portugal}
\affiliation{$^3$European Organization for Nuclear Research (CERN), CH-1211 Geneva 23, Switzerland}
\affiliation{$^4$University of Rochester Laboratory for Laser Energetics, Rochester NY 14623, USA}
\affiliation{$^5$Lawrence Livermore National Laboratory, 7000 East Ave, Livermore, California 94550, USA}
\affiliation{$^6$Rutherford Appleton Laboratory, Chilton, Didcot OX11 0QX, UK}
\affiliation{$^7$Max-Planck-Institut für Kernphysik, Saupfercheckweg 1, D-69117 Heidelberg, Germany}
\affiliation{$^8$Atomic Weapons Establishment, Aldermaston, Reading, Berkshire RG7 4PR, UK}
\affiliation{$^9$Department of Physics, University of Strathclyde, Glasgow G4 0NG, UK}

\date{\today}

\begin{abstract}
A previously unexplored experimental scheme is presented for generating low-divergence, ultra-dense, relativistic, electron-positron beams using 400 GeV/c protons available at facilities such as HiRadMat and AWAKE at CERN. Preliminary Monte-Carlo and Particle-in-cell simulations demonstrate the possibility of generating beams containing $10^{13}-10^{14}$ electron-positron pairs at sufficiently high densities to drive collisionless beam-plasma instabilities, which are expected to play an important role in magnetic field generation and the related radiation signatures of relativistic astrophysical phenomena. The pair beams are quasi-neutral, with size exceeding several skin-depths in all dimensions, allowing for the first time the examination of the effect of competition between transverse and longitudinal instability modes on the growth of magnetic fields. Furthermore, the presented scheme allows for the possibility of controlling the relative density of hadrons to electron-positron pairs in the beam, making it possible to explore the parameter spaces for different astrophysical environments.  

\end{abstract}


\maketitle

\section{Introduction}
The environmental conditions in the magneto-spheres of pulsars, magnetars and black holes are known to present sites of copious electron-positron pair production \cite{goldreich1969pulsar,arons1983pair,blandford1977electromagnetc,begelman1984theory,miller1988active,wardle1998electron}. Outflows from these compact objects, in the form of winds or collimated jets, are inevitably pair-plasma enriched. The energy dissipation mechanisms, that ultimately determine the electromagnetic radiative signatures we measure from Earth, are expected to differ substantially from equivalent electron-ion outflows.  

A particular case of pair-dominated outflows are those thought to generate gamma-ray bursts (GRBs) \cite{meszaros1993relativistic,piran2005physics}. GRBs are among the most luminous events in the universe, yet the precise nature of the emission remains unresolved. It is commonly believed that GRBs result from synchrotron emission of relativistic particles energised at internal shocks \cite{gruzinov2001gamma,chang2008long}. In both the prompt and afterglow GRB emission, it is expected that filamentation-type kinetic beam-plasma instabilities \cite{bludman1960statistical,lee1973electromagnetic,godfrey1975linear} are responsible for the growth of magnetic fields associated with the synchrotron emission \cite{medvedev1999generation, bret2009weibel}, and simulations have suggested that the required field strengths can be amplified in the kind of relativistic collisionless shocks expected to be relevant to GRBs \cite{sakai2004simulation,medvedev2009radiative}. Nevertheless, such studies are constrained by the ability of numerical techniques to fully capture the extreme conditions in GRB outflows. This motivates the development of experimental platforms which can complement simulation studies in exploring the non-linear aspects of beam-plasma instabilities for a range of compositions and densities of beam and background plasmas.

In this paper, we introduce a previously unexplored experimental scheme for generating electron-positron beams using 400 GeV/c protons available at facilities such as HiRadMat \cite{efthymiopoulos2011hiradmat} and AWAKE \cite{gschwendtner2016awake} at CERN. Preliminary Monte-Carlo simulations which model this scheme indicate the possibility of generating low-divergence beams of $10^{13}-10^{14}$ electron-positron pairs with sufficiently high densities to drive filamentation-type beam-plasma instabilities on observable laboratory scales. This number of pairs is significantly higher (by several orders of magnitude) than previously reported laser-produced quasi-neutral pair beams \cite{sarri2015generation,williams2015positron,xu2016ultrashort,warwick2017experimental,williams2020comment}.

\renewcommand{\arraystretch}{1.15}{
\begin{figure*}[t]
\begin{minipage}[b]{0.35\linewidth}
\footnotesize
\begin{tabular}{ccc}
\hline\hline
Parameter & HiRadMat & AWAKE\\
\hline
Beam momentum & $440$ GeV/c & $400$ GeV/c\\
$p^+$ bunch intensity & $1.2\times10^{11}$ & $3\times10^{11}$ \\
Bunch duration & 375 ps & 400 ps \\
1 $\sigma$ beam radius & $0.25-4$ mm & $0.2$ mm\\
\hline\hline
\end{tabular}
\end{minipage}
\begin{minipage}[c]{0.6\linewidth}
\includegraphics[width=1\textwidth]{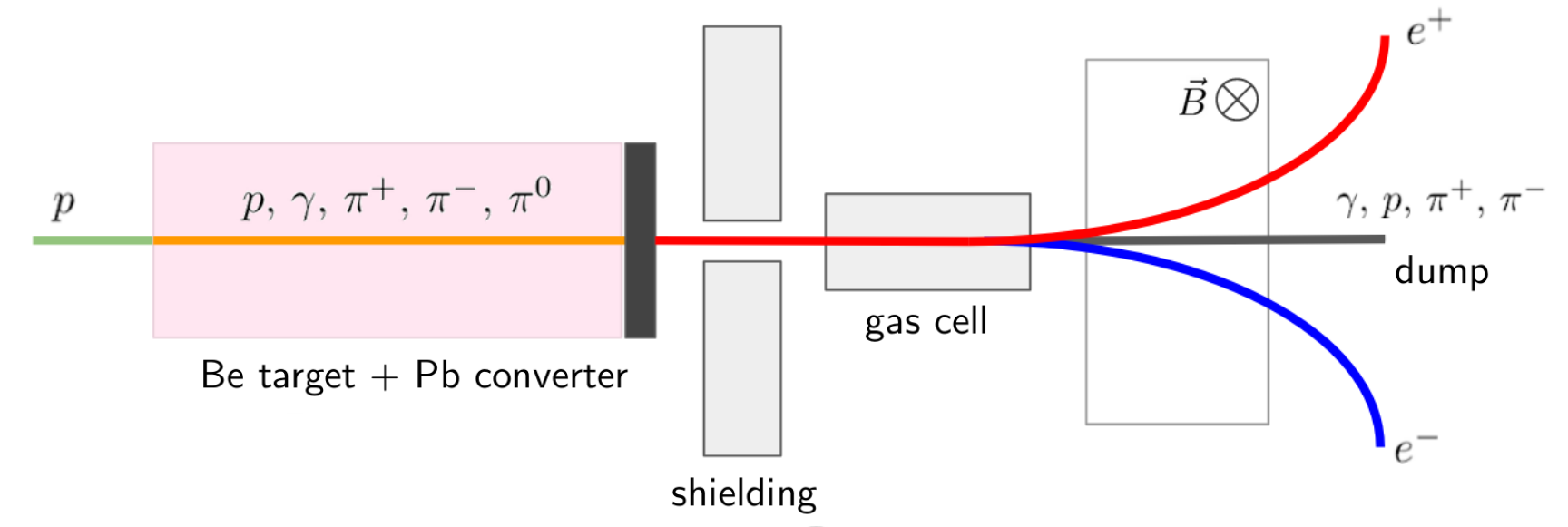}
\end{minipage}
\caption{Left: Beam parameters for 400 GeV/c proton facilities HiRadMat \cite{efthymiopoulos2011hiradmat} and AWAKE \cite{gschwendtner2016awake}. Right: Proposed experimental setup. Beams composed of electrons, positrons, photons, protons and other hadrons are generated using a beryllium target followed by a lead converter. Driving the beam into a gas cell will ionize the gas, forming a background plasma where the beam-plasma interaction can be studied. Since the bulk of the electrons and positrons in the beam have much smaller momentum than the hadrons, dipole magnets can be used to deflect $e^+e^-$ out of the beam to study their energy spectra, while the hadrons are deflected less and are absorbed by the beam dump.}
\label{fig:parameters_setup}
\end{figure*}
}

As well as electron-positron pairs, the beams expected to be generated contain a smaller density of hadrons such as protons and pions. The scheme we introduce allows for the possibility to control the density of these particles relative to electron-positron pairs by several orders of magnitude. This is useful because jet composition remains an important unresolved question surrounding the powering of GRBs. In the fireball model of GRBs, baryon-loading of the jet is discussed as an important parameter in determining the Lorentz factor of the stream \cite{meszaros1993relativistic}, and the role of photoproduction of mesons (and also neutrons and neutrinos) at internal shocks of the fireball model has been discussed as an explanation for anomalous spectral components in the observed prompt emission \cite{asano2009hadronic,asano2009prompt}. The effect of streaming ions on the growth and saturation of magnetic fields via filamentation-like instabilities has been investigated in Particle-in-cell (PIC) simulations \cite{shukla2012enhancement}, and can now be explored experimentally for the first time.

Furthermore, the generated beams have sufficient longitudinal extent to observe obliquely-growing instability modes that are otherwise suppressed in shorter beams \cite{warwick2017experimental}. This is important because obliquely-growing instability modes that compete with transverse current filamentation instability are expected to affect the fraction of bulk kinetic energy of the beam which is converted into magnetic and electric fields, and hence is important for modelling radiative emission processes.

This paper presents simulations and experimental feasibility of the introduced experimental setup. We show preliminary results of Monte-Carlo simulations characterizing the generated electron-positron-hadron beams, and present PIC simulations modelling the propagation of these beams through a background plasma in the laboratory. These simulations demonstrate the development of kinetic instabilities and the growth of feasibly measurable magnetic fields exceeding magnitudes of 0.1 T.

\section{Experimental Scheme}
HiRadMat (High-Radiation to Materials) \cite{efthymiopoulos2011hiradmat} and AWAKE (Advanced Proton Driven Plasma Wakefield Acceleration Experiment) \cite{gschwendtner2016awake} are facilities at CERN which can provide high-intensity 400 GeV/c proton beams up to a maximum intensity of several $10^{11}$ protons per 400 ps pulse. A summary of the beam parameters of these facilities is given in Fig. \ref{fig:parameters_setup}.

In proton-nucleon interactions with centre of mass energy, $\sqrt s$, in excess of GeV, particles produced in hadronic interactions come mostly from the hadronisation of quarks and gluons. A shower of protons, pions, kaons and other hadrons is observed. In particular, a significant component of this shower is a copious number of neutral pions, each of which undergoes electromagnetic decay to two photons on a timescale $\mathcal{O}(10^{-16}$ s) in the $\pi^0$ rest frame. A highly-directional beam of GeV-energy photons is produced in the target, which predominantly loses energy via $e^+e^-$ pair production. The generated $e^+e^-$ lose energy via the generation of bremsstrahlung at a rate approximately proportional to their energy, and so a cascade of copious $e^+e^-$ and $\gamma$ develops. This represents the dominant channel for production of electron-positron pairs initiated with a GeV/c proton beam \cite{charitonidis2020priv}.

For ultra-relativistic streams such as the pair beam systems we generate in this scheme, and those relevant to GRBs (in which Lorentz factors are expected to be in the range $\gamma_b \sim 10^2-10^6$), transverse current filamentation instability (CFI) and oblique instability (OBI) will be the dominant beam-plasma instabilities leading to growth of magnetic fields \cite{bret2009weibel}. The growth of these instabilities is observed when the physical beam size exceeds the background plasma skin depth ($c/\omega_p$), where $\omega_p=\sqrt{4\pi n_e e^2/m_e}$ is the plasma frequency, $n_e$ is the background plasma density, $e$ and $m_e$ are the electron charge and mass, and $c$ is the speed of light. Linear kinetic plasma theory gives estimates of the fastest growth rates of CFI and OBI in the cold distribution function limit \cite{bret2006stabilization} as $\Gamma_{\mathrm{CFI}} \sim  \beta_b\sqrt{\alpha/\gamma_b}~\omega_p$ and $\Gamma_{\mathrm{OBI}} \sim \sqrt{3}/2^{4/3}~(\alpha/\gamma_b)^{1/3}~\omega_p$, where $\beta_b$ and $\gamma_b$ are the Lorentz factors of the beam, and $\alpha$ is the beam-plasma density ratio. Together, these set a requirement that the density of the $e^+e^-$ pair beams must be sufficiently large that collective plasma instabilities grow fast enough to be observed on laboratory timescales, while the background plasma is dense enough for the skin depth to be smaller than the physical beam size.

The easiest way to produce electron-positron pairs from a high-energy photon beam is the conversion process in a high-Z material, such as lead. In a scheme which uses GeV/c protons, preliminary Monte-Carlo simulations indicate that we can increase the maximum pair beam density above what can be obtained by a lead converter alone. We do this by preceding the converter with a beryllium target. Beryllium has a relatively short nucleon interaction length compared to its radiation length, such that a large number of high-energy $\gamma$ can be generated from $\pi^0$ decays in the beryllium with minimal subsequent scattering. This additional flux of photons enhances the densities of pairs that can be generated in the lead converter.

A schematic demonstrating this idea is shown in Fig. \ref{fig:parameters_setup}. To study the growth of beam-plasma instabilities as the generated electron-positron-hadron beams propagate through a background plasma, we can drive the beams immediately into a gas cell. The gas in the cell will ionize (due to both photo-ionization and proton collisions with the neutral atoms) to form a background plasma where beam-plasma instabilities discussed above can be observed. 

\section{Monte-Carlo Simulations}
To better understand the pair beams we create using this scheme, Monte-Carlo modelling was performed using the particle transport code FLUKA \cite{ferrari2005fluka,bohlen2014fluka}, which can accurately simulate hadronic interactions and electromagnetic cascades as a 400 GeV/c proton beam propagates through a solid target of beryllium and lead. As an input for the simulations, we assume a proton beam corresponding to repeatable experimental conditions at HiRadMat. That is, a collimated proton beam with an essentially monochromatic spectral profile peaked at 440 GeV/c (with width corresponding to 0.03\% from the central energy), and a Gaussian transverse beam profile with $\sigma = 0.5$ mm.

Beams of at least $10^5$ protons were simulated to interact with various combinations of thicknesses of beryllium target and lead converter. Beam characteristics such as size, divergence and energy spectra were recorded for the different components of the generated beams escaping the converter rear.

The simulated beams are observed to contain a dominating fluence of $10^{13}-10^{14}$ electron-positron pairs and $\gamma$-rays, along with a smaller number (several tens of times smaller) of protons and other hadron species, the most numerous of these being charged pions.

The transverse radial beam profiles of different beam components are reasonably well-described by a Lorentzian function, and fitting was used to obtain FWHM beam diameters and peak fluences. Estimates of peak volume densities were obtained from peak fluences by assuming a beam duration of 375 ps.

The dependencies of peak component densities on target and converter thicknesses can be found in Fig. 2 for four cases; (a) and (b) show the peak particle densities when beryllium and lead are used on their own, while (c) and (d) show the sensitivities of particle densities to a change in thickness of beryllium or lead from a configuration which gives a high density of $e^+e^-$ pairs (30 cm beryllium target and 4 cm lead converter). As expected, we immediately see that $e^+e^-$ density is much more sensitive to the thickness of the high-Z lead converter. An increase in target thickness will only increase the density of a beam component if additional particle generation is more significant than density decreases due to beam divergence or depletion of particles in processes such as decay, absorption, and annihilation. In all four plots, proton density decreases with target thickness, as protons scatter off the target nuclei. This is more noticeable in the beryllium thickness scans which cover more nuclear collision lengths than the lead scans. Density of electron-positron pairs increases as soon as high-energy photons are generated and cascades initiated, but densities of photons and $e^+e^-$ pairs are higher in lead where the length scale associated with initiation of electromagnetic cascades is much smaller.

\begin{figure}[t]
\centering
\includegraphics[width=\columnwidth]{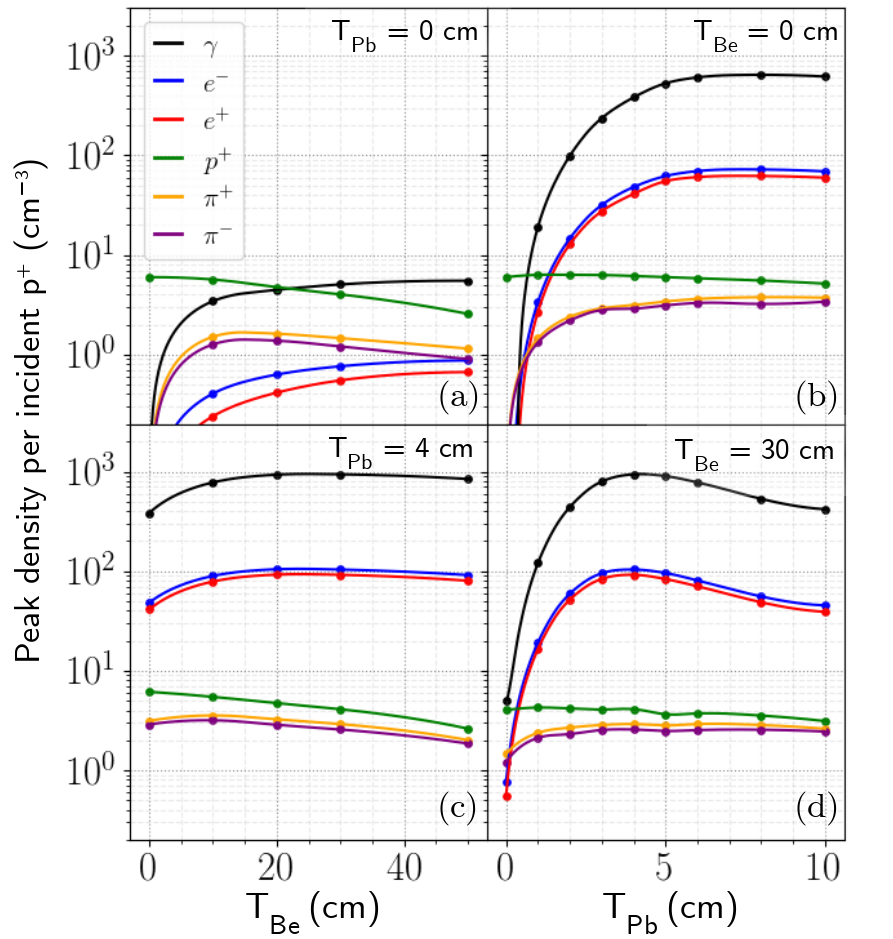}
\label{fig:peak_densities}
\caption{The dependencies of peak densities of each beam species on Be target and Pb converter thicknesses is shown for four configurations. (a) and (b) show the densities obtained for single-component targets of beryllium and lead, while (c) and (d) show the sensitivities of particle densities to a change in thickness of beryllium or lead from a configuration that generates a high density of $e^+e^-$ pairs (that is, 30 cm beryllium target with a 4 cm lead converter). The largest pair beam densities are only achieved by using a configuration that contains both beryllium and lead, and the thickness of lead can be modified to alter the ratio of $e^+e^-$ to hadrons in the beam. Densities are obtained assuming an incident $p^+$ beam with radius $\sigma =$ 0.5 mm, and are presented in units per incident proton, so that the numbers can be scaled to the bunch intensity of the proton facility. A pulse duration of 375 ps is assumed to obtain the peak density from the simulated peak fluence.}
\end{figure}

{
\renewcommand{\arraystretch}{1.25}
\begin{table*}[t]
\caption{Summary of the characteristics of significant particle components of the beam, obtained from Monte-Carlo numerical simulations for the case of a beryllium target and lead converter with thicknesses 30 cm and 4 cm, respectively. All quantities are calculated based on the particles which escape the rear surface of the lead converter. The peak fluence and beam diameter are obtained by fitting the transverse density profile of the escaping beam to a Lorentzian profile. Similarly the beam divergence is the FWHM of the angular distribution of the escaping beam fitted to a Lorentzian profile. The peak fluence is used to infer the peak volume density in the laboratory frame by assuming a 375 ps bunch duration. Yield and peak densities are given per $10^{11}$ incident protons with a beam radius $\sigma =$ 0.5 mm.}
\begin{ruledtabular}
\scriptsize
\begin{tabular}{cccccc}
Species & Yield per $10^{11}$ $p^+$ & Peak density per $10^{11}$ $p^+$ (cm$^{-3}$) & Divergence (mrad) & Beam diameter (mm)  \\
\colrule
$e^-$ & $1.5\times10^{13}$  & $1.0\times10^{13}$ & 25.3 & 2.6  \\
$e^+$ & $1.3\times10^{13}$  & $0.9\times10^{13}$ & 25.3 & 2.6  \\
$p^+$ & $4.0\times10^{11}$  & $4.1\times10^{11}$ & 0.28 & 1.3  \\
$\pi^+$ & $5.5\times10^{11}$  & $2.9\times10^{11}$ & 10.0 & 1.9  \\
$\pi^-$ & $5.6\times10^{11}$ & $2.6\times10^{11}$ & 11.9 & 2.0  \\
\colrule
$\gamma$ & $2.8\times10^{14}$ & $9.4\times10^{13}$ & 17.0 & 3.0 
\end{tabular}
\end{ruledtabular}
\label{tab:characteristics}
\end{table*}
}
\begin{figure*}[t]
\centering
\caption{Energy spectra (left) and angle-position phase space plots (right) obtained in the case of a 30 cm beryllium target and 4 cm lead converter. The simulation setup is the same as the one mentioned in Table \ref{tab:characteristics}. The energy spectra are displayed in the ranges where their spectra are most significant, while insets display the spectra extending to much higher energies. The angle-position phase space plots are normalized and displayed with a colour mapping that clearly depicts the half-maxima.}
\includegraphics[width=0.95\textwidth]{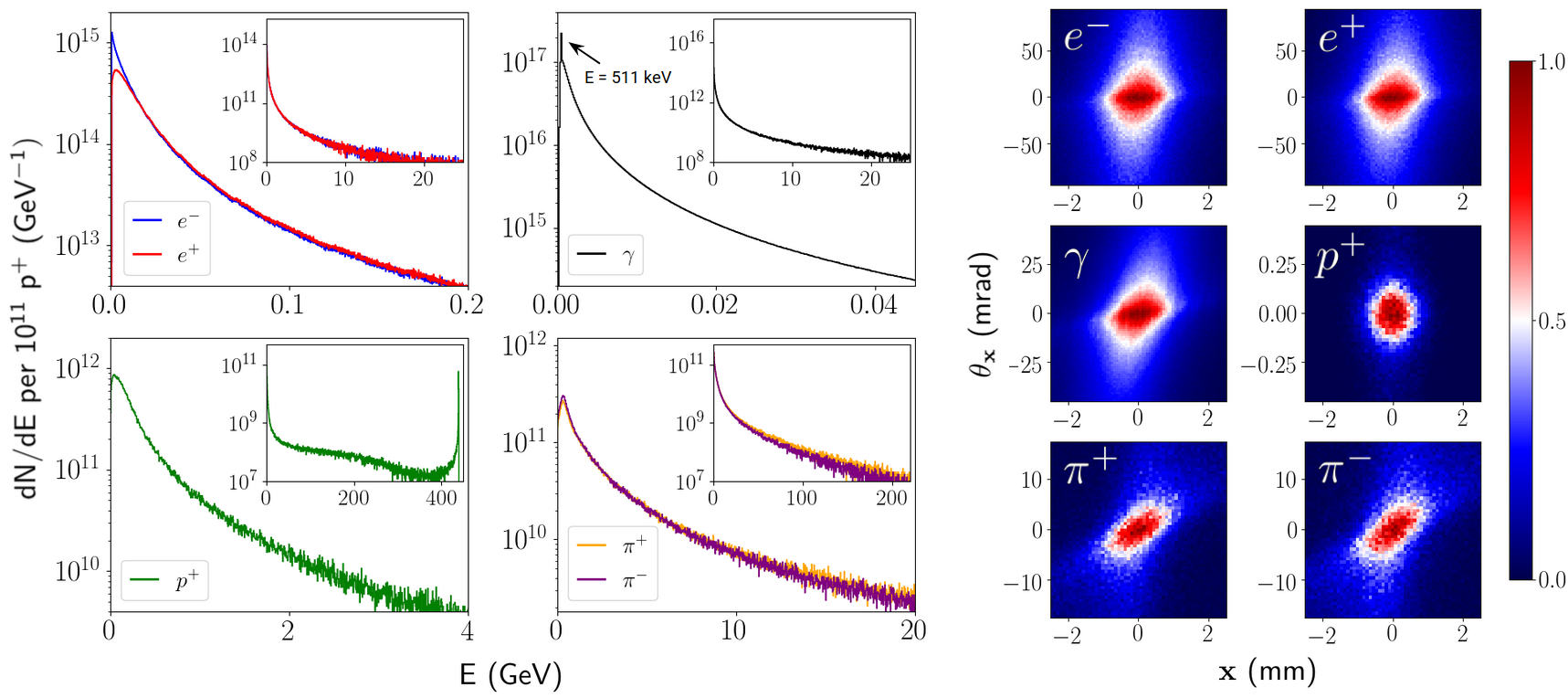}
\label{fig:energy_phase}
\end{figure*}

By comparing the peak densities of pairs in these four plots, we can see that the largest pair beam densities and the largest ratios of pair density to hadron density can only be achieved by using a configuration that contains both beryllium and lead. Using a 30 cm beryllium target and 4 cm lead converter can lead to $e^+e^-$ pair densities in excess of $10^{13}$ cm$^{-3}$ (see column 3 of Table \ref{tab:characteristics}). Higher densities are achievable with higher intensity proton pulses of smaller beam diameter. The thickness of lead can be modified in an experiment to dramatically adjust the ratio of densities of $e^+e^-$ to hadrons, without changing the hadron density significantly. This can allow us to probe jets with a variety of compositions.

For the case of a 30 cm beryllium target and 4 cm lead converter, Fig. \ref{fig:energy_phase} shows the energy spectra and angle-position phase space plots for the significant beam components as they emerge from the rear of the converter. 

The $e^+$ and $e^-$ spectra are very similar, differing only at energies less than 10 MeV, where the annihilation cross section of positrons becomes significant. The spectra are dominated by particles which have only 10's of MeV, but extend up to tens of GeV in their high-energy tails. These characteristics are matched by the photon spectra, with the addition of a spectral peak at 511 keV resulting from electron-positron annihilation.

\begin{figure*}[t]
\centering
\includegraphics[width=0.93\textwidth]{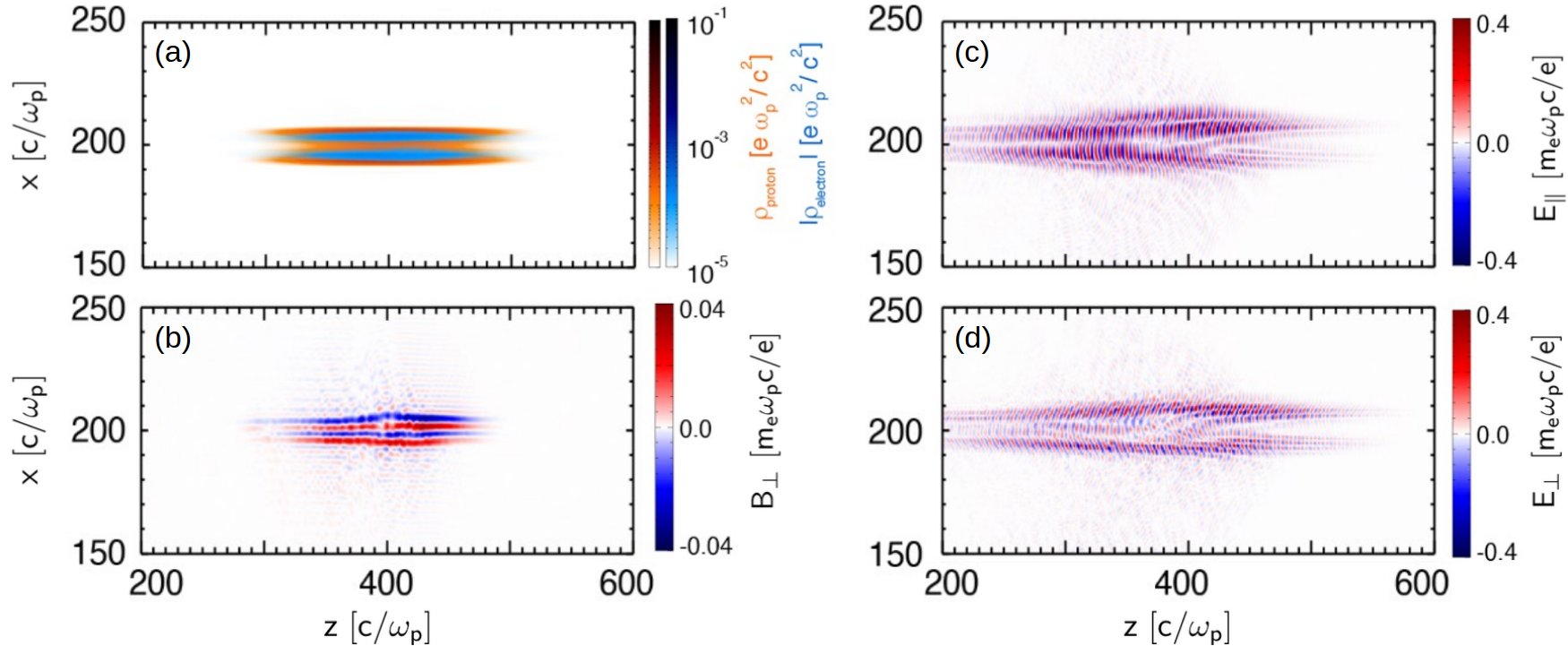}
\caption{Simulation results of the interaction between an electron-positron-proton bunch and a static plasma with density $10^{14}~\mathrm{cm}^{-3}$ at a time $t=705~[1/\omega_p]=1.27~\mathrm{ns}$. (a) Density filaments of electrons (blue) and protons (red). (b) Transverse magnetic fields ($B_{\bot}$) filaments due to current filamentation. (c) Longitudinal electric fields ($E_{\parallel}$), and (d) transverse electric fields ($E_{\bot}$) attributed to space charge and inductive effects. Units are such that one plasma period [$1/\omega_p$] corresponds to $[1/\omega_p]=1.8~\mathrm{ps}$, while one skin depth [$c/\omega_p$] corresponds to $[c/\omega_p]=530~\mathrm{\mu m}$, and magnetic and electric field units are $[m_e\omega_pc/e]=3.2~\mathrm{T}$ and $[m_e\omega_pc/e]=\mathrm{GV}~\mathrm{m}^{-1}$ respectively.}
\label{fig:pic_sims}
\end{figure*}

The proton spectra appears bimodal in distribution. We see a peak at 440 GeV corresponding to the protons in the initial beam which have not lost a significant fraction of their energy, and we also see much lower energy protons resulting from the inelastic hadronic scattering. Spectra with similar characteristics are seen for the charged pions, except with the omission of the high-energy contribution from an initial beam.

We can get an idea of the extent to which high beam densities are maintained as the beam propagates by looking at its overall divergence (the distribution of the angles between the beam axis and the particle trajectory of all the particles of a species as they exit the rear of the converter). For $e^+e^-$ and $\gamma$ we observe that the emerging beams have an overall divergence of $15-30$ mrad (see column 4 of Table \ref{tab:characteristics}). The pions have divergence of 10 mrad, which means the ratio of beam density of pions to $e^+e^-$ will not change much as the beam propagates. However, the lower divergence of the protons due to the directionality of the high energy component of the beam, means that care must be taken to position the gas cell close to the converter rear in order to maximise the dominance of $e^+e^-$ pair density over proton density. 

For all beam components we find the beam diameter to be mm-scale (see column 5 of Table \ref{tab:characteristics}), meaning the background plasma must exceed densities of  $10^{13}-10^{14}~\mathrm{cm}^{-3}$ for the transverse beam size to be larger than the plasma skin depth and for us to observe filamentation instabilities. Referring back to the theoretical growth rates of CFI and OBI, we find that with these background plasma densities, instability growth rates on the order of picoseconds might be expected, given mean beam Lorentz factors $\gamma_b \sim 100$ and beam densities $10^{13}-10^{14}~\mathrm{cm}^{-3}$. We also find that CFI and OBI have closely competing growth rates.

For beams such as ours, which are relativistically hot ($k_B T > m_ec^2$), the true scalings and growth rates will be different from the cold beam approximations. So we have performed PIC simulations to better understand the non-linear growth of CFI and OBI in the case of our $e^-e^+p^+$ beams propagating through a background plasma. From PIC simulations we can gain better insight into the competition between CFI and OBI growth, and better estimates of the timescales for instability growth and saturation. We can also obtain the magnitude of the energy expected to be converted into magnetic and electric fields, and whether the generated fields are of a sufficiently high magnitude that they might be measurable in an experiment.

\section{Particle-in-cell simulations}
The fully relativistic, massively parallel, PIC code OSIRIS \cite{fonseca2002osiris,fonseca2008one}, which has been used extensively to model relativistic beam-plasma interactions \cite{silva2003interpenetrating,shukla2018conditions,shukla2020interaction}, was used to perform two-dimensional PIC simulations of an electron-positron-proton bunch through a background plasma. The simulations used a moving window travelling at $c$, with absorbing boundary conditions, and dimensions $800\times400~(c/\omega_p)^2$ divided into $8000\times4000$ cells with $6\times 6$ particles per cell for plasma electrons and beam particles. Electron-positron-proton beams were initialized at the entrance of a stationary plasma with density profile given by $n_b = n_{b0}\exp(-r^2/\sigma_{r}^2 - z^2/\sigma_z^2)$, where the bunch peak densities are $n_{b0}=10^{13}~\mathrm{cm}^{-3}$ for electrons and positrons, and $n_{b0} \sim 10^{11}~\mathrm{cm}^{-3}$ for the protons. The bunch length and transverse waist are $\sigma_{z}=7.0~\mathrm{cm}=132~c/\omega_p$ and $\sigma_r=0.15~\mathrm{cm}=2.8~c/\omega_p$ respectively, with the skin depth of the background plasma $c/\omega_p$ corresponding to a plasma density $n_{p}=10^{14}~\mathrm{cm}^{-3}$. An ion-electron mass ratio $m_p/m_e=1836$ is used.

Studies which have investigated the effect of temperature on growth of filamentation and oblique modes have found that growth rates are strongly suppressed in beams with large transverse temperature \cite{bret2010multidimensional}, but are less dependent on large longitudinal energy spreads \cite{shukla2018conditions}. Therefore, we set each beam component to propagate along the z-axis with Lorentz factor set by the mean longitudinal momentum (without a thermal momentum spread), while a thermal momentum spread is included in the transverse direction that corresponds to the mean transverse momentum. For the electrons and positrons, we choose a bulk Lorentz factor of $\gamma \sim \langle p_\parallel \rangle/m_ec \sim 100$, and transverse momentum spread corresponding to $k_B T_\bot \sim \langle p_\bot\rangle c \sim 7.5~\mathrm{MeV}$, while $\gamma \sim \langle p_\parallel \rangle/m_pc \sim 50$ and $k_B T_\bot \sim \langle p_\bot\rangle c \sim 370~\mathrm{MeV}$ for the protons.

\begin{figure}[t]
\centering
\includegraphics[width=0.95\columnwidth]{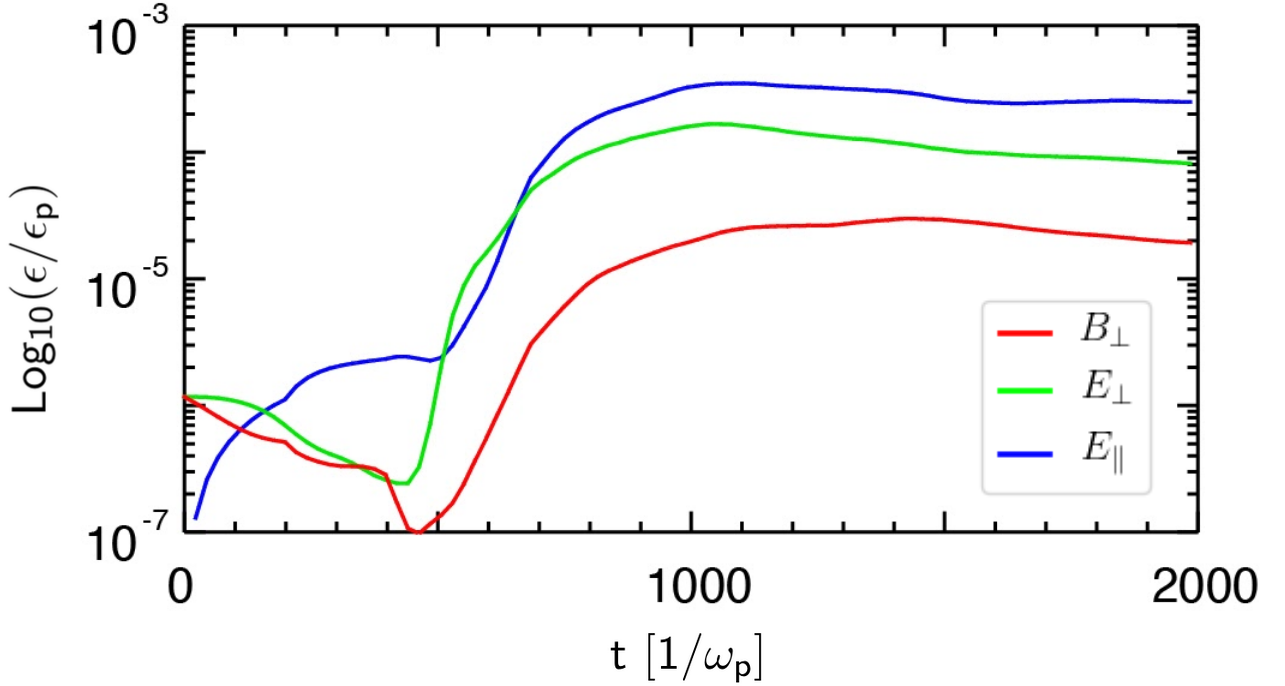}
\label{fig:field_energies}
\caption{Evolution of energies contained within transverse magnetic field $\epsilon_{B\bot}$ (red), transverse electric field $\epsilon_{E\bot}$ (green) and longitudinal electric field $\epsilon_{E\parallel}$ (blue) as the beam propagates, normalized to the initial kinetic energy of the beam $\epsilon_p=(\gamma_b-1)V$ where $V$ is the volume of the beam.}
\end{figure}

Simulation results are illustrated in Fig. \ref{fig:pic_sims}, showing the spatial-temporal evolution of protons and electrons in the beam (Fig. \ref{fig:pic_sims}a), the formation of transverse magnetic filaments (Fig. \ref{fig:pic_sims}b), and the typical electric field structure (Fig. \ref{fig:pic_sims}c and \ref{fig:pic_sims}d). In our simulations we observe the breaking-up of the beam into current filaments (with width on the order of $c/\omega_p \sim 500~\mathrm{\mu m}$) as the electron-positron-proton bunch enters into the background plasma. Any charge separation in the beam generates micro-currents which reinforces initial perturbations, causing the growth of electromagnetic plasma instability and the exponential growth of electromagnetic fields. This is demonstrated in Fig. 5, where the temporal evolution of the magnetic and electric field energies is shown as a function of time, normalized to the initial bulk kinetic energy of beam.

The fields saturate after $t=1000~[1/\omega_p]=1.8~\mathrm{ns}$, and magnetic fields with amplitudes of 0.13 T are generated via the current filamentation instability (corresponding to $\sim10^{-5}$ of the initial beam energy). Simultaneously, longitudinal and transverse electric fields are observed with magnitudes exceeding $300~\mathrm{MV}~\mathrm{m^{-1}}$, which can be attributed to space charge and inductive effects \cite{shukla2018conditions}. This is contrary to the case of electron driven plasma wakefields, where no filamentation of the beam is observed, nor the generation of strong magnetic fields \cite{chen1985acceleration,chen1987plasma,lotov2007acceleration}. The protons in the beam are not found to play a significant role in the collective plasma dynamics, although they do contribute to space charge effects.

Our simulations show the emergence of oblique modes and tilted filamentation, which reduces the growth of magnetic fields. This is seen in previous similar simulation studies \cite{shukla2018conditions}. Importantly, the effect of the growth of oblique modes on magnetic field generation can only be studied with beams that have a large enough longitudinal extent to allow coupling between transverse and longitudinal beam instability modes. It cannot be studied using quasi-neutral $e^+e^-$ beams generated at laser facilities \cite{sarri2015generation,williams2015positron,xu2016ultrashort,warwick2017experimental,williams2020comment}, which are typically limited by having maximum beam durations of approximately $50~\mathrm{fs}$.

\section{Summary}
The growth of kinetic plasma instabilities of relevance to astrophysical phenomena such as GRBs has been investigated for a previously unexplored experimental platform in which low-divergence, ultra-dense, quasi-neutral, electron-positron pair beams are generated using 400 GeV/c protons available at facilities such as AWAKE and HiRadMat at CERN.

Monte-Carlo simulations demonstrate the possibility of generating beams that contain $10^{13}-10^{14}$ electron-positron pairs along with a smaller number ($10^{11}-10^{12}$) of hadrons such as protons and pions. Particle-in-cell simulations have shown that beams interacting with a background plasma will exhibit collective plasma effects and the generation of magnetic fields exceeding 0.1 T via filamentation instabilities, which saturate after 50 cm of beam propagation.

This platform represents a significant step forward for experiments aimed at exploring relativistic pair-plasma phenomena in a laboratory setting. We have demonstrated the experimental feasibility of isolating and studying the non-linear evolution of several key instabilities that are presently limited to numerical experiments, allowing for the possibility of observing the effects of obliquely-growing filamentation modes and the role of hadrons on magnetic field generation in the development of kinetic plasma instabilities. By changing experimental parameters such as incident proton density, target thickness and density of background plasma, different configurations corresponding to fireballs traversing an overdense or underdense background medium can be examined, and the composition of different astrophysical scenarios can be studied. 

\vspace{0.4cm}
The research leading to these results has received funding from AWE plc., the Central Laser Facility (Science and Technology Facilities Council of the United Kingdom), the European Research Council (through grant ERC-2015-AdG, GA No. 695088), Fundação para a Ciência e a Tecnologia (through grant EXPL/FIS-PLA/0834/2012), and the Department of Energy Office of Fusion Energy (DE-SC0017950). FLUKA simulations were performed using the Science and Technology Facilities Council Scientific Computing Department's SCARF cluster, and OSIRIS simulations were performed at the Marconi-Broadwell (CINECA, Italy).

\bibliography{references}

\end{document}